\documentclass[12pt,a4paper]{article}
\usepackage{amsmath}
\usepackage[dvips]{epsfig}
\usepackage{feynmf}             
\title{
  {\vspace{-2cm} \normalsize
     \epsfig{figure=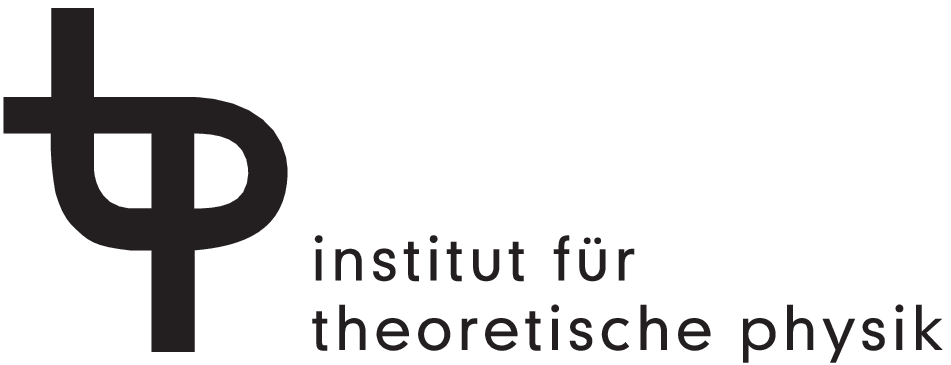,width=80mm}
     \hfill\parbox[b][30mm][t]{35mm}{MS-TP-03-6 \\
                                     DESY 03-080 \\
                                     hep-th/0307053}  }\\[25mm]
Perturbation theory for the two-dimensional abelian Higgs model
in the unitary gauge
}
\author{Gernot M\"unster and Enno E.\ Scholz%
        \thanks{present address: DESY, Notkestr.~85, D-22603 Hamburg}\\
        Institut f\"ur Theoretische Physik,
        Universit\"at M\"unster\\
        Wilhelm-Klemm-Str.~9, D-48149 M\"unster, Germany\\
        e-mail: munsteg@uni-muenster.de, enno.e.scholz@desy.de}
\date{August 08, 2003}
\newcommand{\E}{\mathrm{e}}
\newcommand{\I}{\mathrm{i}}
\begin{document}
\maketitle

\begin{abstract}
In the unitary gauge the unphysical degrees of freedom of spontaneously
broken gauge theories are eliminated. The Feynman rules are simpler than
in other gauges, but it is non-renormalizable by the rules of power
counting. On the other hand, it is formally equal to the limit $\xi \to
0$ of the renormalizable R$_{\xi}$-gauge. We consider perturbation
theory to one-loop order in the R$_{\xi}$-gauge and in the unitary gauge
for the case of the two-dimensional abelian Higgs model. An apparent
conflict between the unitary gauge and the limit $\xi \to 0$ of the
R$_{\xi}$-gauge is resolved, and it is demonstrated that results for
physical quantities can be obtained in the unitary gauge.
\end{abstract}

\begin{fmffile}{mfhiggs}
%
\section{Introduction}

For theories of interacting gauge and Higgs fields with spontaneously
broken gauge symmetry two well known gauges are the unitary gauge
(U-gauge) \cite{Weinberg1} and the renormalizable R$_{\xi}$-gauge
\cite{FLS,Abers}.  In the U-gauge the gauge-variant transversal part of
the Higgs field has been eliminated and the Lagrangian only contains
physical degrees of freedom. Although the Feynman rules in the U-gauge
are simpler, it is usually not used for perturbative calculations.  The
reason for this is the fact that for large momenta the gauge field
propagator grows faster than in the R$_{\xi}$-gauge.  Consequently the
model in the U-gauge appears to be unrenormalizable by the usual
power-counting rules.  In the R$_{\xi}$-gauge, on the other hand, more
fields have to be taken into account, namely the unphysical components
of the Higgs field and the ghost field.  The Feynman rules are more
complicated and there are more diagrams to be calculated.  The
advantage is that the model in the R$_{\xi}$-gauge is manifestly
renormalizable.

With the help of Slavnov-Taylor identities it can formally be shown
that renormalized on-shell quantities are independent of the gauge
\cite{LeeZJ2,LeeZJ4,Abers}. Such physical quantities should therefore
in principle be calculable in the U-gauge.  It appears, however, that
the cancellation of divergent terms is a delicate matter.  In practice,
calculations in the unitary gauge have sometimes led to results, which
are in conflict with those obtained in other gauges
\cite{Dolan,Weinberg2}.

In this paper we address the possibility of doing perturbation theory
in the U-gauge and the relation between the R$_{\xi}$-gauge and the
U-gauge.  For simplicity we restrict ourselves to the two-dimensional
abelian Higgs model.  It contains all the features we would like to
discuss, but the explicit calculations are easier than in non-abelian
models in 4 dimensions.  We perform the perturbative calculations of
off-shell quantities on the one-loop level.  Ultraviolet divergencies
are treated by means of dimensional regularization, where the number of
dimensions of space-time are taken to be $D=2-2\epsilon$.

Formally, the U-gauge is obtained by taking the limit $\xi \to 0$ of
the R$_{\xi}$-gauge.  Applying this prescription naively, results are
obtained, which do not coincide with those of the U-gauge.  We discuss
the origin of this discrepancy, which is related to the fact that the
limits $\epsilon \to 0$ and $\xi \to 0$ are not interchangeable, and
discuss how to go from the R$_{\xi}$-gauge to the U-gauge properly.

We would like to point out that renormalization of the 4-dimensional
abelian Higgs model in the unitary gauge has been discussed by Sonoda
\cite{Sonoda} with the help of suitable choice of interpolating fields.

%
\section{The two-dimensional abelian Higgs model}

The model contains a real vector field $A_{\mu}(x)$ and a complex scalar
Higgs field $\phi(x)$.  We shall consider the theory in a
two-dimensional space-time with a Euclidean metric.  The Lagrangian is
\begin{equation}
\mathcal{L} = \frac{1}{4} F_{\mu\nu} F_{\mu\nu}
+ |D_{\mu} \phi|^2 + V(\phi),
\end{equation}
where
\begin{equation}
F_{\mu\nu} = \partial_{\mu} A_{\nu} - \partial_{\nu} A_{\mu}\,,
\end{equation}
\begin{equation}
D_{\mu} = \partial_{\mu} - \I e A_{\mu}\,,
\end{equation}
\begin{equation}
V(\phi) = - \frac{m^2}{2} |\phi|^2 + \frac{g}{6} |\phi|^4\,,
\end{equation}
and $e$ and $g$ are coupling constants.  The potential is of the mexican
hat type with its minima at
\begin{equation}
|\phi| = \frac{v}{\sqrt{2}}\,, \quad \mbox{where} \quad
v^2 = \frac{3 m^2}{g}.
\end{equation}
The Lagrangian is invariant under local gauge transformations
\begin{eqnarray}
\phi(x) &\rightarrow& \phi'(x) = \E^{-\I \alpha(x)} \phi(x)\,,\\
A_{\mu}(x) &\rightarrow& A'_{\mu}(x) =
A_{\mu}(x) - \frac{1}{e} \partial_{\mu} \alpha(x)\,.
\end{eqnarray}

%
\subsection{Unitary gauge}

The scalar field can be written in the form
\begin{equation}
\phi(x) = \rho(x) \, \E^{\I \omega(x)}
\end{equation}
with real $\rho(x)$ and $\omega(x)$. The U-gauge is obtained by choosing
the gauge transformation function as $\alpha(x) = \omega(x)$. The
transformed fields are then
\begin{eqnarray}
\phi'(x) &=& \rho(x),\\
A'_{\mu}(x) &=& A_{\mu}(x) - \frac{1}{e} \partial_{\mu} \omega(x)
\doteq B_{\mu}(x),\\
F'_{\mu\nu} &=& \partial_{\mu} B_{\nu} - \partial_{\nu} B_{\mu}\,.
\end{eqnarray}
They represent the gauge invariant physical degrees of freedom. In terms
of $\rho(x)$ the potential can be expressed as
\begin{equation}
V(\rho) = \frac{g}{6} \left( \rho^2 - \frac{v^2}{2} \right)^2
+ \mbox{const.}
\end{equation}
After expanding the scalar field around the minimum of the potential as
\begin{equation}
\rho(x) = \frac{1}{\sqrt{2}} ( v + \sigma(x) ),
\end{equation}
the Lagrangian reads, up to an irrelevant constant,
\begin{eqnarray}
\mathcal{L} &=&
\frac{1}{4} F_{\mu\nu} F_{\mu\nu}
+ \frac{1}{2} e^2 v^2 B_{\mu}^2
+ \frac{1}{2} (\partial_{\mu} \sigma)^2
+ \frac{m^2}{2} \sigma^2 \nonumber\\
&& + \frac{gv}{3!} \sigma^3
+ \frac{g}{4!} \sigma^4
+ e^2 v \sigma B_{\mu}^2
+ \frac{1}{2} e^2 \sigma^2 B_{\mu}^2\,.
\end{eqnarray}
One can read off that the Higgs scalar $\sigma$ has mass $m$ and the
vector field $B_{\mu}$ is massive with mass $m_{v}= e v$. From the
Lagrangian the following Feynman rules are obtained.
\begin{itemize}
\item
scalar propagator:
$\Delta_{\sigma}(k) = (m^2 + k^2)^{-1}
= \ \raisebox{-3\unitlength}{%
\begin{fmfgraph}(60,10)
\fmfleft{i}
\fmfright{o}
\fmf{plain}{i,o}
\end{fmfgraph}
}$
\item
gauge field propagator:
$\Delta_{\mu\nu}(k) = (m^2_{v} + k^2)^{-1} \left( \delta_{\mu\nu}
+ \frac{k_{\mu} k_{\nu}}{m^2_{v}} \right)
=\;\;\;\;\; \raisebox{-7\unitlength}{%
\begin{fmfgraph*}(60,20)
\fmfstraight
\fmfleft{i}
\fmfright{o}
\fmf{boson}{i,o}
\fmflabel{$\mu$}{i}
\fmflabel{$\nu$}{o}
\end{fmfgraph*}
}$
\item
$\sigma^3$-vertex: $-gv
= \raisebox{-25\unitlength}{%
\begin{fmfgraph}(50,50)
\fmfstraight
\fmftop{i}
\fmfbottomn{o}{2}
\fmf{plain}{i,v}
\fmf{plain}{o1,v}
\fmf{plain}{o2,v}
\fmfdot{v}
\end{fmfgraph}
}$
\item
$\sigma^4$-vertex: $-g
= \raisebox{-25\unitlength}{%
\begin{fmfgraph}(50,50)
\fmfleftn{i}{2}
\fmfrightn{o}{2}
\fmf{plain}{i1,v}
\fmf{plain}{i2,v}
\fmf{plain}{o1,v}
\fmf{plain}{o2,v}
\fmfdot{v}
\end{fmfgraph}
}$
\item
$\sigma - B_{\mu}^2$-vertex:
$-2e^2v\delta_{\mu\nu}
= \raisebox{-25\unitlength}{%
\begin{fmfgraph*}(50,50)
\fmfstraight
\fmftop{i}
\fmfbottomn{o}{2}
\fmf{plain}{i,v}
\fmf{boson}{o1,v}
\fmf{boson}{o2,v}
\fmfdot{v}
\fmflabel{$\mu$}{o1}
\fmflabel{$\nu$}{o2}
\end{fmfgraph*}
}$
\vspace*{1cm}
\item
$\sigma^2 - B_{\mu}^2$-vertex:
$-2e^2\delta_{\mu\nu}
= \raisebox{-25\unitlength}{%
\begin{fmfgraph*}(50,50)
\fmfstraight
\fmftopn{i}{2}
\fmfbottomn{o}{2}
\fmf{plain}{i1,v}
\fmf{plain}{i2,v}
\fmf{boson}{o1,v}
\fmf{boson}{o2,v}
\fmflabel{$\mu$}{o1}
\fmflabel{$\nu$}{o2}
\fmfdot{v}
\end{fmfgraph*}
}$
\vspace*{1cm}
\end{itemize}
With these Feynman rules one can write down expressions for various
Green functions. In order not to overlook subtleties, it should be taken
into account, however, that the functional integral measure for the
scalar field $\sigma(x)$ is not the standard one. For each point $x$ the
measure is, up to a constant factor,
\begin{equation}
d(\mathrm{Re}\,\phi(x))\,d(\mathrm{Im}\,\phi(x)) =
\rho(x) d\rho(x) d\omega(x)\,.
\end{equation}
The functional integral measure for the scalar field is therefore
\begin{equation}
\prod_{x} (v + \sigma(x))\,d\sigma(x)
\doteq \det J \prod_{x} d\sigma(x)
\end{equation}
with \cite{Gerstein}
\begin{equation}
J(x,y) = \delta(x-y) \left( (v + \sigma(x)) \right).
\end{equation}
One can try to argue that $\det J$ does not affect the perturbative
results, at least in dimensional regularization \cite{Appel}. But it is
safer to keep this term for the moment. With the help of ghost fields
we can write
\begin{equation}
\det J = \int\!\!\mathcal{D}c \mathcal{D}\bar{c}\ \E^{-S_{gh}}\,,
\end{equation}
with
\begin{equation}
S_{gh} = e^2 v \int\!\!dx\ \bar{c}(x) \left( (v + \sigma(x)) \right)
c(x)\,.
\end{equation}
The prefactor $e^2 v$ has been chosen such that comparison with similar
terms in the R$_{\xi}$-gauge is facilitated. The Lagrangian gets the
additional ghost terms
\begin{equation}
m^2_{v}\ \bar{c} c + e^2 v\,\sigma\,\bar{c} c
\end{equation}
and the Feynman rules are augmented by
\begin{itemize}
\item
ghost propagator: $(m^2_{v})^{-1}
= \ \raisebox{-12\unitlength}{%
\begin{fmfgraph}(50,30)
\fmfleft{l}
\fmfright{r}
\fmf{dots_arrow}{l,r}
\end{fmfgraph}
}$
\item
ghost-$\sigma$-vertex: $- e^2 v
= \raisebox{-25\unitlength}{%
\begin{fmfgraph}(50,50)
\fmftop{t}
\fmfbottom{b1,b2}
\fmf{plain}{t,i}
\fmf{dots_arrow}{b1,i,b2}
\fmfdot{i}
\end{fmfgraph}
}$
\end{itemize}
In the one-loop order the Green's functions of the scalar and vector
fields get additional contributions from ghost loops. These are
proportional to
\begin{equation}
\int\!\frac{d^D k}{(2\pi)^D}\ \frac{1}{m^2_{v}}\,.
\end{equation}
In dimensional regularization these contributions vanish due to the rule
\cite{Leibb}
\begin{equation}
\label{dimrule}
\int\!\frac{d^D k}{(2\pi)^D}\ (k^2)^{\alpha} = 0\,, \quad \mathrm{for}\
\alpha \ge 0\,.
\end{equation}
This justifies one neglecting the measure factor $\det J$. The ghost
fields introduced above are, however, useful in the discussion of the
relation between the U-gauge and the R$_{\xi}$-gauge.

%
\subsection{R$_{\xi}$-gauge}

In the R$_{\xi}$-gauge the Higgs field is decomposed into its real and
imaginary parts as
\begin{equation}
\phi(x) = \frac{1}{\sqrt{2}} ( v + \phi_{1}(x) + \I \phi_{2}(x) )\,.
\end{equation}
Expanding the Lagrangian in terms of these fields, a mixing term between
$A_{\mu}(x)$ and $\phi_{2}(x)$ appears on the quadratic level. The
R$_{\xi}$-gauge is specified by the gauge fixing function
\begin{equation}
F = \partial_{\mu} A_{\mu} - \frac{e v}{\xi}\,\phi_{2}\,,
\end{equation}
where $\xi > 0$ is a real parameter. The gauge fixing term to be added
to the Lagrangian is
\begin{equation}
\mathcal{L}_{gf} = \frac{\xi}{2} F^2\,.
\end{equation}
It eliminates the $A_{\mu}(x) - \phi_{2}(x)$ mixing term. The gauge
fixing procedure yields the Faddeev-Popov determinant $\det M_F$, where
\begin{equation}
M_F = - \partial_{\mu}^2 + \frac{e^2 v}{\xi} (v + \phi_{1})\,.
\end{equation}
As usual it can be represented in terms of ghost fields via a ghost
Lagrangian
\begin{equation}
\mathcal{L}_{gh} = \xi\,\bar{c}(x) M_F\, c(x)
= - \xi\,\bar{c}(x) \partial_{\mu}^2 c(x)
+ m^2_{v}\,\bar{c}(x) c(x) + e^2 v\,\phi_{1}(x) \bar{c}(x) c(x)\,.
\end{equation}
By suitable normalization of the ghost fields the prefactor $\xi$ has
been chosen for later convenience. In contrast to the case of QED the
ghost term cannot be neglected since it contains an interaction between
the Higgs and the ghost fields.

{}From the total Lagrangian the following Feynman rules are derived. In
order to save space, the graphical representations are shown only for
new types of propagators or vertices.
\begin{itemize}
\item
$\phi_{1}$ propagator: $\Delta_{\phi_{1}}(k) = (m^2 + k^2)^{-1}$
\item
$\phi_{2}$ propagator:
$\Delta_{\phi_{2}}(k) = (\frac{m^2_{v}}{\xi} + k^2)^{-1}
= \ \raisebox{-7\unitlength}{%
\begin{fmfgraph}(60,20)
\fmfstraight
\fmfleft{i}
\fmfright{o}
\fmf{dashes}{i,o}
\end{fmfgraph}
}$
\item
gauge field propagator:
$\Delta_{\mu\nu,\,\xi}(k) =
(m^2_{v} + k^2)^{-1} \left( \delta_{\mu\nu}
- \frac{k_{\mu} k_{\nu}}{k^2} \right)
+\frac{1}{\xi}\,(\frac{m^2_{v}}{\xi} + k^2)^{-1}
\frac{k_{\mu} k_{\nu}}{k^2}$
\item
ghost propagator: $\Delta_{c}(k) = (m^2_{v} + \xi k^2)^{-1}$
\item
$\phi_{1}^3$-vertex: $- gv$
\item
$\phi_{1}^4$-vertex: $- g$
\item
$\phi_{1} \phi_{2}^2$-vertex: $- \frac{gv}{3}$
\item
$\phi_{1}^2 \phi_{2}^2$-vertex: $- \frac{g}{3}$
\item
$\phi_{2}^4$-vertex: $- g$
\item
$A_{\mu}^2 \phi_{1}$-vertex: $- 2 e^2 v \delta_{\mu\nu}$
\item
$A_{\mu} \phi_{1} \phi_{2}$-vertex: $- \I e (k_1 - k_2)_{\mu}\,
=\quad \raisebox{-25\unitlength}{%
\begin{fmfgraph*}(50,50)
\fmfstraight
\fmfbottom{b}
\fmftopn{p}{2}
\fmf{boson}{b,v}
\fmf{plain,label=$k_1$,label.side=right}{p1,v}
\fmf{dashes,label=$k_2$,label.side=left}{p2,v}
\fmfdot{v}
\fmflabel{$\mu$}{b}
\end{fmfgraph*}
}$
\item
$A_{\mu}^2 \phi_{1}^2$-vertex: $- 2 e^2 \delta_{\mu\nu}$
\item
$A_{\mu}^2 \phi_{2}^2$-vertex: $- 2 e^2 \delta_{\mu\nu}$
\item
$\phi_{1} \bar{c} c$-vertex: $- e^2 v$
\end{itemize}
Comparing with the Feynman rules of the U-gauge we observe that in the
limit $\xi \to 0$ the propagators and vertices involving the fields
$\phi_{1}$, $A_{\mu}$ and $\bar{c}, c$ go over to those of the fields
$\sigma$, $B_{\mu}$ and $\bar{c}, c$ in the U-gauge. Moreover the
$\phi_{2}$ propagator
\begin{equation}
\Delta_{\phi_{2}}(k) = \frac{\xi}{m^2_{v} + \xi k^2}
\stackrel{\xi \to 0}{\longrightarrow} 0
\end{equation}
vanishes in this limit. In this sense the U-gauge formally corresponds
to the $\xi \to 0$ limit of the R$_{\xi}$-gauge \cite{Abers,Appel}.
Indeed, in this limit the gauge fixing function forces the imaginary
component $\phi_{2}$ of the scalar field to vanish, which corresponds
to the U-gauge.

Two other special cases are known in the literature. The limit $\xi \to
\infty$ yields the Landau gauge, in which the vector propagator is
purely transversal. The case $\xi=1$ is the Feynman gauge, which has the
simplest Feynman rules.

%
\section{Perturbation theory}

In this section we consider Green's functions of the abelian Higgs
model in perturbation theory in one-loop order. For the treatment of
divergencies we employ dimensional regularization with
$D = 2 - 2 \epsilon$ dimensions. The coupling constants are replaced
by
\begin{eqnarray}
e &\rightarrow& \mu^{\epsilon} e\,, \nonumber\\
g &\rightarrow& \mu^{2 \epsilon} g\,, \nonumber\\
v &\rightarrow& \mu^{- \epsilon} v\,,
\end{eqnarray}
where $\mu$ is an arbitrary mass scale.

The one-loop corrections are of order $g$ or $e^2$ relative to the tree
level terms. As usual, fractions $e^2 / g$ are counted as being of
order 1. Two-loop and higher corrections are of order $g^2$, $g e^2$ or
$e^4$.

%
\subsection{Scalar propagator}

Let us start with the scalar propagator. We write its inverse as
\begin{equation}
G^{-1}(p) = m^2 + p^2 + \Sigma(p)\,,
\end{equation}
where the self-energy $\Sigma(p)$ is given by the sum of one-particle
irreducible, amputated propagator diagrams. For the $\sigma$-propagator
in the U-gauge we obtain
\begin{eqnarray}
\nonumber-\Sigma_{\sigma}(p^2)&=&\Bigg(%
\begin{fmfgraph*}(60,30)
\fmfstraight
\fmftop{o}
\fmfbottom{l,v1,r}
\fmf{plain,label=$p$}{l,v1}
\fmf{plain,right=1}{v1,o}
\fmf{plain,left=1}{v1,o}
\fmf{plain}{v1,r}
\fmfdot{v1}
\end{fmfgraph*}%
\,+\,
\begin{fmfgraph*}(60,30)
\fmfstraight
\fmftop{o}
\fmfbottom{l,v1,r}
\fmf{plain,label=$p$}{l,v1}
\fmf{boson,right}{v1,o}
\fmf{boson,right}{o,v1}
\fmf{plain}{v1,r}
\fmfdot{v1}
\end{fmfgraph*}%
\,+\,
\begin{fmfgraph*}(60,30)
\fmfstraight
\fmfbottom{l,v1,r}
\fmftop{o}
\fmf{plain,label=$p$}{l,v1}
\fmf{plain}{v1,r}
\fmf{plain,tension=1.5}{v1,v2}
\fmf{phantom}{v2,i}
\fmf{phantom}{i,o}
\fmf{plain,left,tension=0.5}{v2,o}
\fmf{plain,left,tension=0.5}{o,v2}
\fmfdotn{v}{2}
\end{fmfgraph*}%
\,+\,
\begin{fmfgraph*}(60,30)
\fmfstraight
\fmfbottom{l,v1,r}
\fmftop{o}
\fmf{plain,label=$p$}{l,v1}
\fmf{plain}{v1,r}
\fmf{plain,tension=1.5}{v1,v2}
\fmf{phantom}{v2,i}
\fmf{phantom}{i,o}
\fmf{boson,left,tension=0.5}{v2,o}
\fmf{boson,left,tension=0.5}{o,v2}
\fmfdotn{v}{2}
\end{fmfgraph*}%
\\
\nonumber\\
\nonumber&&\,+\,\raisebox{-15\unitlength}{%
\begin{fmfgraph*}(60,30)
\fmfstraight
\fmfleft{l}
\fmfright{r}
\fmf{plain,label=$p$}{l,v1}
\fmf{plain,left,tension=0.5}{v1,v2}
\fmf{plain,left,tension=0.5}{v2,v1}
\fmf{plain}{v2,r}
\fmfdotn{v}{2}
\end{fmfgraph*}}%
\,+\,
\raisebox{-15\unitlength}{%
\begin{fmfgraph*}(60,30)
\fmfstraight
\fmfleft{l}
\fmfright{r}
\fmf{plain,label=$p$}{l,v1}
\fmf{boson,left,tension=0.3}{v1,v2}
\fmf{boson,left,tension=0.3}{v2,v1}
\fmf{plain}{v2,r}
\fmfdotn{v}{2}
\end{fmfgraph*}}%
\Bigg)_{\textrm{\scriptsize{amp}}}\\
\nonumber\\
\nonumber&=&\frac{1}{4\pi} \Bigg\{\,g \bigg[
\left( 1-\frac{p^2}{3m^2} \right)
\left(\frac{1}{\epsilon}-\gamma-\ln\frac{m^2}{4\pi\mu^2}\right)
\:+\:\frac{p^2}{3m^2}\ln\frac{3e^2}{g}\\
\nonumber&&\:+\frac{3m^2}{2\sqrt{p^4+4p^2m^2}}
\ln\frac{p^2+2m^2+\sqrt{p^4+4p^2m^2}}{p^2+2m^2-\sqrt{p^4+4p^2m^2}}
\,\bigg]\\
\nonumber&&+e^2\bigg[2\left(\frac{1}{\epsilon}-\gamma-2
-\ln\frac{m_{v}^2}{4\pi\mu^2}\right)\\
&&\:+\frac{(p^2+2m_{v}^2)^2}{2m_{v}^2\sqrt{p^4+4p^2m_{v}^2}}
\ln\frac{p^2+2m_{v}^2+\sqrt{p^4+4p^2m_{v}^2}}
{p^2+2m_{v}^2-\sqrt{p^4+4p^2m_{v}^2}}\,\bigg]\Bigg\},
\end{eqnarray}
where $\gamma = 0.57721\ldots$ is Euler's constant.

{}From the propagator the renormalized mass and the wave function
renormalization constant can be obtained. We shall consider two schemes
here. In the first scheme the renormalized mass $m_R$ and
renormalization constant $Z_R$ are defined by
\begin{equation}
G_{\sigma}^{-1}(p) = \frac{1}{Z_R} \{ m^2_R + p^2 +
\mathcal{O}(p^4)\}\,,
\end{equation}
which amounts to
\begin{eqnarray}
Z_R^{-1} &=& 1 +
\left. \frac{\partial \Sigma_{\sigma}}{\partial p^2}\right|_{p=0} \\
m^2_R &=& Z_R ( m^2 + \Sigma_{\sigma}(0) )\,.
\end{eqnarray}
This gives
\begin{eqnarray}
\nonumber m_R^2&=& m^2 - \frac{g}{12\pi}
\bigg[\frac{43}{12}\,+\,4\left(\frac{1}{\epsilon}-\gamma
-\ln\frac{m^2}{4\pi\mu^2}\right)
\,-\,\ln\frac{3e^2}{g}\bigg]\\
&&-\:\frac{e^2}{2\pi}\left[\frac{1}{\epsilon}-\gamma-1
-\ln\frac{m_{v}^2}{4\pi\mu^2}\right],\\
Z_R&=&1 + \frac{g}{4\pi m^2}
\bigg[\frac{11}{36}-\frac{1}{3}\left(\frac{1}{\epsilon}-\gamma
-\ln\frac{m_{v}^2}{4\pi\mu^2}\right)\bigg].
\end{eqnarray}

In the second scheme the renormalized mass is taken to be the physical
mass $m_{\sigma}$, given by the pole of the propagator,
\begin{equation}
G_{\sigma}^{-1}\left( (\I m_{\sigma}, 0) \right) = 0\,,
\end{equation}
and the wave function renormalization constant $Z_{\sigma}$ is the
corresponding residuum,
\begin{equation}
G_{\sigma}(p) \simeq \frac{Z_{\sigma}}{p^2 + m_{\sigma}^2}
\quad \mathrm{for} \quad p^2 \to - m_{\sigma}^2\,.
\end{equation}
We get
\begin{eqnarray}
\nonumber m_\sigma^2&=&m^2\:-\:\frac{g}{4\pi}
\Bigg[\frac{4}{3}\left(\frac{1}{\epsilon}-\gamma
-\ln\frac{m^2}{4\pi\mu^2}\right)
\:-\:\frac{1}{3}\ln\frac{3e^2}{g}
\:+\:2\sqrt{3}\,\textrm{arccot}\sqrt{3}\\
&&\:+\frac{2}{3}\,
\frac{\left(1-6\frac{e^2}{g}\right)^2}{\sqrt{12\frac{e^2}{g}-1}}\,
\textrm{arccot}\sqrt{12\frac{e^2}{g}-1}\,\Bigg]
-\frac{e^2}{2\pi}
\bigg[\frac{1}{\epsilon}-\gamma-2-\ln\frac{m_{v}^2}{4\pi\mu^2}\bigg],
\end{eqnarray}
\begin{eqnarray}
\nonumber Z_\sigma&=&1\:+\:\frac{g}{4\pi m^2}
\Bigg[-1\:+\:\frac{2\sqrt{3}}{3}\,\textrm{arccot}\sqrt{3}
\:-\:\frac{1}{3}\left(\frac{1}{\epsilon}-\gamma
-\ln\frac{m_{v}^2}{4\pi\mu^2}\right)\\
&&\:+\frac{\left(1-6\frac{e^2}{g}\right)^2}
{3\left(1-12\frac{e^2}{g}\right)}
\:+\:\frac{2}{3}\,\frac{\left(1-6\frac{e^2}{g}\right)
\left(1-12\frac{e^2}{g}-36\frac{e^4}{g^2}\right)}
{\left(12\frac{e^2}{g}-1\right)^{\frac{3}{2}}}\,
\textrm{arccot}\sqrt{12\frac{e^2}{g}-1}\,\Bigg],
\end{eqnarray}
where we assume $g \leq 12 e^2$ for the analytic continuation.

The corresponding propagator in the R$_{\xi}$-gauge is the
$\phi_{1}$-propagator. Its self-energy in one-loop order is
\begin{eqnarray}
\nonumber-\Sigma_{\phi_1,\textrm{\scriptsize{R$_{\xi}$}}}(p^2)&=&\Bigg(%
\begin{fmfgraph*}(60,30)
\fmfstraight
\fmftop{o}
\fmfbottom{l,v1,r}
\fmf{plain,label=$p$}{l,v1}
\fmf{plain,right=1}{v1,o}
\fmf{plain,left=1}{v1,o}
\fmf{plain}{v1,r}
\fmfdot{v1}
\end{fmfgraph*}%
\,+\,
\begin{fmfgraph*}(60,30)
\fmfstraight
\fmftop{o}
\fmfbottom{l,v1,r}
\fmf{plain,label=$p$}{l,v1}
\fmf{dashes,right=1}{v1,o}
\fmf{dashes,left=1}{v1,o}
\fmf{plain}{v1,r}
\fmfdot{v1}
\end{fmfgraph*}%
\,+\,
\begin{fmfgraph*}(60,30)
\fmfstraight
\fmftop{o}
\fmfbottom{l,v1,r}
\fmf{plain,label=$p$}{l,v1}
\fmf{boson,right}{v1,o}
\fmf{boson,right}{o,v1}
\fmf{plain}{v1,r}
\fmfdot{v1}
\end{fmfgraph*}%
\,+\,
\begin{fmfgraph*}(60,30)
\fmfstraight
\fmfbottom{l,v1,r}
\fmftop{o}
\fmf{plain,label=$p$}{l,v1}
\fmf{plain}{v1,r}
\fmf{plain,tension=1.5}{v1,v2}
\fmf{phantom}{v2,i}
\fmf{phantom}{i,o}
\fmf{plain,left,tension=0.5}{v2,o}
\fmf{plain,left,tension=0.5}{o,v2}
\fmfdotn{v}{2}
\end{fmfgraph*}\\
\nonumber&&+\,
\begin{fmfgraph*}(60,30)
\fmfstraight
\fmfbottom{l,v1,r}
\fmftop{o}
\fmf{plain,label=$p$}{l,v1}
\fmf{plain}{v1,r}
\fmf{plain,tension=1.5}{v1,v2}
\fmf{phantom}{v2,i}
\fmf{phantom}{i,o}
\fmf{dashes,left,tension=0.5}{v2,o}
\fmf{dashes,left,tension=0.5}{o,v2}
\fmfdotn{v}{2}
\end{fmfgraph*}%
\,+\,
\begin{fmfgraph*}(60,30)
\fmfstraight
\fmfbottom{l,v1,r}
\fmftop{o}
\fmf{plain,label=$p$}{l,v1}
\fmf{plain}{v1,r}
\fmf{plain,tension=1.5}{v1,v2}
\fmf{phantom}{v2,i}
\fmf{phantom}{i,o}
\fmf{boson,left,tension=0.5}{v2,o}
\fmf{boson,left,tension=0.5}{o,v2}
\fmfdotn{v}{2}
\end{fmfgraph*}%
\,+\,
\begin{fmfgraph*}(60,30)
\fmfstraight
\fmfbottom{l,v1,r}
\fmftop{o}
\fmf{plain,label=$p$}{l,v1}
\fmf{plain}{v1,r}
\fmf{plain,tension=1.5}{v1,v2}
\fmf{phantom}{v2,i}
\fmf{phantom}{i,o}
\fmf{dots_arrow,left,tension=0.5}{v2,o}
\fmf{dots,left=1,tension=0.5}{o,v2}
\fmfdotn{v}{2}
\end{fmfgraph*}%
\,+\,\raisebox{-15\unitlength}{%
\begin{fmfgraph*}(60,30)
\fmfstraight
\fmfleft{l}
\fmfright{r}
\fmf{plain,label=$p$}{l,v1}
\fmf{plain,left,tension=0.5}{v1,v2}
\fmf{plain,left,tension=0.5}{v2,v1}
\fmf{plain}{v2,r}
\fmfdotn{v}{2}
\end{fmfgraph*}}\\
\nonumber&&+\,\raisebox{-15\unitlength}{%
\begin{fmfgraph*}(60,30)
\fmfstraight
\fmfleft{l}
\fmfright{r}
\fmf{plain,label=$p$}{l,v1}
\fmf{dashes,left,tension=0.5}{v1,v2}
\fmf{dashes,left,tension=0.5}{v2,v1}
\fmf{plain}{v2,r}
\fmfdotn{v}{2}
\end{fmfgraph*}}%
\,+\,\raisebox{-15\unitlength}{%
\begin{fmfgraph*}(60,30)
\fmfstraight
\fmfleft{l}
\fmfright{r}
\fmf{plain,label=$p$}{l,v1}
\fmf{dots_arrow,left,tension=0.5}{v1,v2}
\fmf{dots_arrow,left,tension=0.5}{v2,v1}
\fmf{plain}{v2,r}
\fmfdotn{v}{2}
\end{fmfgraph*}}%
\,+\,\raisebox{-15\unitlength}{%
\begin{fmfgraph*}(60,30)
\fmfstraight
\fmfleft{l}
\fmfright{r}
\fmf{plain,label=$p$}{l,v1}
\fmf{dashes,left,tension=0.4}{v1,v2}
\fmf{boson,left,tension=0.4}{v2,v1}
\fmf{plain}{v2,r}
\fmfdotn{v}{2}
\end{fmfgraph*}}%
\,+\,\raisebox{-15\unitlength}{%
\begin{fmfgraph*}(60,30)
\fmfstraight
\fmfleft{l}
\fmfright{r}
\fmf{plain,label=$p$}{l,v1}
\fmf{boson,left,tension=0.4}{v1,v2}
\fmf{boson,left,tension=0.4}{v2,v1}
\fmf{plain}{v2,r}
\fmfdotn{v}{2}
\end{fmfgraph*}}%
\Bigg)_{\textrm{\scriptsize{amp}}}\\
\nonumber\\
\nonumber&=&\frac{1}{4\pi}\Bigg\{\,g\Bigg[\frac{4}{3}
\left(\frac{1}{\epsilon}-\gamma-\ln\frac{m^2}{4\pi\mu^2}\right)
\:-\:\frac{1}{3}\ln\frac{3e^2}{g}\:+\:\frac{p^2+m^2}{3m^2}\ln\xi\\
\nonumber&&\:+\frac{3m^2}{2\sqrt{p^4+4m^2p^2}}
\ln\frac{p^2+2m^2+\sqrt{p^4+4m^2p^2}}{p^2+2m^2-\sqrt{p^4+4m^2p^2}}\\
\nonumber&&\:+\frac{m^4-p^4}{6m^2\sqrt{p^4+4p^2\frac{m_{v}^2}{\xi}}}
\ln\frac{p^2+2\frac{m_{v}^2}{\xi}+\sqrt{p^4+4p^2\frac{m_{v}^2}{\xi}}}
{p^2+2\frac{m_{v}^2}{\xi}-\sqrt{p^4+4p^2\frac{m_{v}^2}{\xi}}}\,\Bigg]\\
\nonumber&&+e^2\bigg[2\left(\frac{1}{\epsilon}-\gamma-2
-\ln\frac{m_{v}^2}{4\pi\mu^2}\right)\\
&&\:+\frac{(p^2+2m_{v}^2)^2}{2m_{v}^2\sqrt{p^4+4p^2m_{v}^2}}
\ln\frac{p^2+2m_{v}^2+\sqrt{p^4+4p^2m_{v}^2}}
{p^2+2m_{v}^2-\sqrt{p^4+4p^2m_{v}^2}}\,\bigg]\Bigg\}.
\end{eqnarray}
This expression is valid if
$p^2 \geq -4 m^2$, $p^2 \geq -4 m^2_{v}$ and
$p^2 \geq -4 m^2_{v} / \xi$ are fulfilled.
For the renormalized masses and renormalization constants we obtain
\begin{eqnarray}
\nonumber m_{R,\textrm{\scriptsize{R$_{\xi}$}}}^2&=&m^2
\:-\:\frac{g}{12\pi}\bigg[\frac{43}{12}\:+\:4\left(\frac{1}{\epsilon}
-\gamma-\ln\frac{m^2}{4\pi\mu^2}\right)
\:-\:\ln\frac{3e^2}{g}\:+\:\xi\frac{g}{6e^2}
\:+\:\xi^2\frac{g^2}{108e^4}\bigg]\\
&&\:-\frac{e^2}{2\pi}\left(\frac{1}{\epsilon}-\gamma-1
-\ln\frac{m_{v}^2}{4\pi\mu^2}\right),\\
Z_{R,\xi}&=&1\;+\;\frac{g}{4\pi m^2}
\left[\frac{11}{36}\:+\:\frac{1}{3}\ln\xi
\:-\:\xi^2\frac{g^2}{324e^4}\right]
\end{eqnarray}
in the first scheme, and

\begin{eqnarray}
\nonumber
m^2_{\phi_{1}}&=&m^2\:-\:\frac{g}{4\pi}\Bigg[\frac{4}{3}
\left(\frac{1}{\epsilon}-\gamma-\ln\frac{m_{v}^2}{4\pi\mu^2}\right)
-\frac{1}{3}\ln\frac{3e^2}{g}+2\sqrt{3}\,\textrm{arccot}\sqrt{3}\\
&&+\frac{2}{3}\,
\frac{\left(1-6\frac{e^2}{g}\right)^2}{\sqrt{12\frac{e^2}{g}-1}}\,
\textrm{arccot}\sqrt{12\frac{e^2}{g}-1}\Bigg]
-\frac{e^2}{2\pi}\bigg[\left(\frac{1}{\epsilon}-\gamma-2
-\ln\frac{m_{v}^2}{4\pi\mu^2}\right)\bigg],\\
\nonumber Z_{\phi_{1}}&=&1\;+\;\frac{g}{4\pi m^2}
\Bigg[-1\:+\:\frac{2\sqrt{3}}3\,\textrm{arccot}\sqrt{3}
\:+\:\frac{1}{3}\ln\xi
\:+\:\frac{\left(1-6\frac{e^2}{g}\right)^2}
{3\left(1-12\frac{e^2}{g}\right)}\\
\nonumber&&\:+\:\frac{2}{3}\,\frac{\left(1-6\frac{e^2}{g}\right)
\left(1-12\frac{e^2}{g}-36\frac{e^4}{g^2}\right)}
{\left(12\frac{e^2}{g}-1\right)^{\frac{3}{2}}}\,
\textrm{arccot}\sqrt{12\frac{e^2}{g}-1}
\:+\frac{4}{3\sqrt{12\frac{e^2}{\xi g}-1}}\,
\textrm{arccot}\sqrt{12\frac{e^2}{\xi g}-1}\,\Bigg]\\
\end{eqnarray}
in the second scheme.

In all cases the renormalized propagator
\begin{equation}
G_R (p) = Z^{-1} \, G(p)\,,
\end{equation}
expressed in terms of the renormalized mass, is finite. In the
R$_{\xi}$-gauge it depends on the gauge parameter $\xi$. The
renormalized mass $m_{R,\xi}$, not being a physical on-shell quantity,
also depends on $\xi$. In contrast, the physical mass $m_{\phi_{1}}$ is
independent of $\xi$ as expected.

%
\subsection{The U-gauge limit}

Let us consider the relation between the two gauges. We should expect
that the expressions calculated in the R$_{\xi}$-gauge go over to those
in the U-gauge, if we let $\xi \to 0$. Indeed, for the masses we see
that
\begin{equation}
\lim_{\xi \to 0} m_{R, \xi} = m_R
\end{equation}
in the first scheme, and
\begin{equation}
m_{\phi_{1}} = m_{\sigma}
\end{equation}
in the second scheme.

For the renormalization constants, however, the situation is different.
Both $Z_{R,\,\xi}$ and $Z_{\phi_{1}}$ contain a ($\log \xi$)-term and
appear to diverge as $\xi \to 0$. Here the formal equivalence between
the R$_{\xi}$-gauge in the limit $\xi \to 0$ and the U-gauge seems to
break down.

Let us consider this discrepancy more carefully. The propagator gets
contributions from ghost and $\phi_{2}$-loops, which are of the form
\begin{equation}
I = \mu^{2\epsilon}
\int\!\frac{d^D k}{(2\pi)^D}\ \frac{1}{k^2 + \frac{m^2_{v}}{\xi}}\,.
\end{equation}
In dimensional regularization this is
\begin{equation}
I = \frac{1}{4\pi}
\left( \frac{m^2_{v}}{4\pi \mu^2 \xi} \right)^{- \epsilon}
\Gamma( \epsilon )\,.
\end{equation}
If expanded for small $\epsilon$ in the usual way, it reads
\begin{equation}
I = \frac{1}{4\pi}
\left( \frac{1}{\epsilon} - \gamma - \log \frac{m^2_{v}}{4\pi \mu^2}
+ \log \xi + \mathcal{O}(\epsilon) \right)
\end{equation}
and we find the disturbing ($\log \xi$)-term. This expansion for small
$\epsilon$ is, however, only applicable for fixed finite $\xi$. The
$\xi$ dependence of $I$ is contained in the factor
\begin{equation}
\frac{1}{\epsilon}\ \xi^{\epsilon} =
\frac{1}{\epsilon} + \log \xi + \mathcal{O}(\epsilon)\,.
\end{equation}
If the limit $\xi \to 0$ is taken first, with a positive $\epsilon$,
one gets instead
\begin{equation}
I \stackrel{\xi \to 0}{\longrightarrow} 0\,.
\end{equation}
Alternatively, this can be obtained by writing
\begin{equation}
I = \xi \, \mu^{2\epsilon}
\int\!\frac{d^D k}{(2\pi)^D}\ \frac{1}{\xi k^2 + m^2_{v}}
\end{equation}
and using the rule Eq.~(\ref{dimrule}).

For the other terms involving gauge field loops the integrals are more
complicated, but a detailed analysis shows that similar considerations
hold.

We conclude that the limits $\epsilon \to 0$ and $\xi \to 0$ cannot
be interchanged.  As a consequence, the Laurent expansion in
$\epsilon$ is not compatible with the limit $\xi \to 0$.  In order to
arrive at the U-gauge as a limit of the R$_{\xi}$-gauge, the limit has
to be taken for fixed non-vanishing $\epsilon$ before the resulting
expressions are expanded around $\epsilon = 0$.

In general the small $\xi$- and $\epsilon$-dependence of a diagram
in $D$ dimensions is of the type $\xi^{\alpha ( D_0 - D)}$. The number
$D$ of dimensions has then to be chosen sufficiently small, $D < D_0$,
when taking the limit $\xi \to 0$. In the example above we have
$\alpha=\frac{1}{2}$, $D_0 = 2$.

Taking these considerations into account, the limit $\xi \to 0$ can be
taken for the self-energy, and the resulting expression coincides with
the one in the U-gauge.  Consequently the renormalized masses and
renormalization constants also coincide in this limit.

%
\subsection{Gauge field propagator}

The inverse gauge field propagator can be decomposed into a transversal
and a longitudinal part as
\begin{equation}
G^{-1}_{\mu\nu}(p) =
\left[ \delta_{\mu\nu} - \frac{p_{\mu} p_{\nu}}{p^2} \right]
\left[ m^2_{v} + p^2 + \Pi_1 (p^2) \right]
+ \frac{p_{\mu} p_{\nu}}{p^2}
\left[ m^2_{v} + \xi p^2 + \Pi_2 (p^2) \right].
\end{equation}
In the R$_{\xi}$-gauge the diagrams
\begin{eqnarray}
\nonumber
\begin{fmfgraph*}(60,30)
\fmfstraight
\fmftop{o}
\fmfbottom{l,v1,r}
\fmf{boson,label=$p$}{l,v1}
\fmf{plain,right=1}{v1,o}
\fmf{plain,left=1}{v1,o}
\fmf{boson}{v1,r}
\fmfdot{v1}
\fmflabel{$\mu$}{l}
\fmflabel{$\nu$}{r}
\end{fmfgraph*}%
\hspace*{1cm}%
\begin{fmfgraph*}(60,30)
\fmfstraight
\fmftop{o}
\fmfbottom{l,v1,r}
\fmf{boson,label=$p$}{l,v1}
\fmf{dashes,right=1}{v1,o}
\fmf{dashes,left=1}{v1,o}
\fmf{boson}{v1,r}
\fmfdot{v1}
\fmflabel{$\mu$}{l}
\fmflabel{$\nu$}{r}
\end{fmfgraph*}%
\hspace*{1cm}\raisebox{-15\unitlength}{%
\begin{fmfgraph*}(60,30)
\fmfstraight
\fmfleft{l}
\fmfright{r}
\fmf{boson,label=$p$}{l,v1}
\fmf{plain,left,tension=0.4}{v1,v2}
\fmf{boson,left,tension=0.4}{v2,v1}
\fmf{boson}{v2,r}
\fmfdotn{v}{2}
\fmflabel{$\mu$}{l}
\fmflabel{$\nu$}{r}
\end{fmfgraph*}}%
\hspace*{1cm}\raisebox{-15\unitlength}{%
\begin{fmfgraph*}(60,30)
\fmfstraight
\fmfleft{l}
\fmfright{r}
\fmf{boson,label=$p$}{l,v1}
\fmf{plain,left,tension=0.7}{v1,v2}
\fmf{dashes,left,tension=0.7}{v2,v1}
\fmf{boson}{v2,r}
\fmfdotn{v}{2}
\fmflabel{$\mu$}{l}
\fmflabel{$\nu$}{r}
\end{fmfgraph*}}\\
\begin{fmfgraph*}(60,30)
\fmfstraight
\fmfbottom{l,v1,r}
\fmftop{o}
\fmf{boson,label=$p$}{l,v1}
\fmf{boson}{v1,r}
\fmf{plain,tension=1.5}{v1,v2}
\fmf{phantom}{v2,i}
\fmf{phantom}{i,o}
\fmf{plain,left,tension=0.5}{v2,o}
\fmf{plain,left,tension=0.5}{o,v2}
\fmfdotn{v}{2}
\fmflabel{$\mu$}{l}
\fmflabel{$\nu$}{r}
\end{fmfgraph*}%
\hspace*{1cm}%
\begin{fmfgraph*}(60,30)
\fmfstraight
\fmfbottom{l,v1,r}
\fmftop{o}
\fmf{boson,label=$p$}{l,v1}
\fmf{boson}{v1,r}
\fmf{plain,tension=1.5}{v1,v2}
\fmf{phantom}{v2,i}
\fmf{phantom}{i,o}
\fmf{dashes,left,tension=0.5}{v2,o}
\fmf{dashes,left,tension=0.5}{o,v2}
\fmfdotn{v}{2}
\fmflabel{$\mu$}{l}
\fmflabel{$\nu$}{r}
\end{fmfgraph*}%
\hspace*{1cm}%
\begin{fmfgraph*}(60,30)
\fmfstraight
\fmfbottom{l,v1,r}
\fmftop{o}
\fmf{boson,label=$p$}{l,v1}
\fmf{boson}{v1,r}
\fmf{plain,tension=1.5}{v1,v2}
\fmf{phantom}{v2,i}
\fmf{phantom}{i,o}
\fmf{boson,left,tension=0.5}{v2,o}
\fmf{boson,left,tension=0.5}{o,v2}
\fmfdotn{v}{2}
\fmflabel{$\mu$}{l}
\fmflabel{$\nu$}{r}
\end{fmfgraph*}%
\hspace*{1cm}%
\begin{fmfgraph*}(60,30)
\fmfstraight
\fmfbottom{l,v1,r}
\fmftop{o}
\fmf{boson,label=$p$}{l,v1}
\fmf{boson}{v1,r}
\fmf{plain,tension=1.5}{v1,v2}
\fmf{phantom}{v2,i}
\fmf{phantom}{i,o}
\fmf{dots_arrow,left,tension=0.5}{v2,o}
\fmf{dots,left=1,tension=0.5}{o,v2}
\fmfdotn{v}{2}
\fmflabel{$\mu$}{l}
\fmflabel{$\nu$}{r}
\end{fmfgraph*}
\end{eqnarray}
yield
\begin{eqnarray}
\nonumber\Pi_1 (p^2)&=&-\frac{e^2}{4\pi}
\bigg[4\left(\frac{1}{\epsilon}-\gamma+1-\ln\frac{m^2}{4\pi\mu^2}\right)
\:-\:\frac{p^2+m^2-m_{v}^2}{p^2}\ln\frac{3e^2}{g}\\
\nonumber&&\:-\frac{(p^2+m^2-m_{v}^2)^2}
{p^2\sqrt{(p^2+m^2-m_{v}^2)^2+4p^2m_{v}^2}}
\ln\frac{p^2+m^2+m_{v}^2+\sqrt{(p^2+m^2-m_{v}^2)^2+4p^2m_{v}^2}}
{p^2+m^2+m_{v}^2-\sqrt{(p^2+m^2-m_{v}^2)^2+4p^2m_{v}^2}}\\
&&\:+\frac{6e^2}{g}\left(\frac{1}{\epsilon}-\gamma-2
-\ln\frac{m_{v}^2}{4\pi\mu^2}\right)\,\bigg],\\
\nonumber\Pi_{2,\xi} (p^2)&=&-\frac{e^2}{4\pi}
\Bigg[4\left(\frac{1}{\epsilon}-\gamma-\ln\frac{m^2}{4\pi\mu^2}\right)
\:+\:\frac{6e^2}{g}\left(\frac{1}{\epsilon}-\gamma-2
-\ln\frac{m_{v}^2}{4\pi\mu^2}\right)\:+\:2\ln\xi\\
\nonumber&&-\frac{p^2-m^2+m_{v}^2}{p^2}\ln\frac{3e^2}{g}\\
\nonumber&&+\frac{\sqrt{(p^2+m^2-m_{v}^2)^2+4p^2m_{v}^2}}{p^2}
\ln\frac{p^2+m^2+m_{v}^2+\sqrt{(p^2+m^2-m_{v}^2)^2+4p^2m_{v}^2}}
{p^2+m^2+m_{v}^2-\sqrt{(p^2+m^2-m_{v}^2)^2+4p^2m_{v}^2}}\\
\nonumber&&-\frac{p^2+2m^2-2\frac{m_{v}^2}{\xi}}
{\sqrt{\left(p^2+m^2-\frac{m_{v}^2}{\xi}\right)^2
+4p^2\frac{m_{v}^2}{\xi}}}
\ln\frac{p^2+m^2+\frac{m_{v}^2}{\xi}
+\sqrt{\left(p^2+m^2-\frac{m_{v}^2}{\xi}\right)^2
+4p^2\frac{m_{v}^2}{\xi}}}
{p^2+m^2+\frac{m_{v}^2}{\xi}
-\sqrt{\left(p^2+m^2-\frac{m_{v}^2}{\xi}\right)^2
+4p^2\frac{m_{v}^2}{\xi}}}\,\Bigg].\\
\end{eqnarray}

The transversal part is manifestly independent of the gauge parameter
$\xi$ and is identical to the one in the U-gauge. This is generally
true, as has been discussed in \cite{LeeZJ3,Haeus} to all orders in
perturbation theory. The renormalized vector mass and corresponding
renormalization factor are derived from the transversal propagator and
are equal, too, in both gauges. One obtains
\begin{eqnarray}
\nonumber m^2_{R,v}&=&m_{v}^2\:-\:\frac{e^2}{2\pi}
\left[2\left(\frac{1}{\epsilon}-\gamma-\ln\frac{m^2}{4\pi\mu^2}\right)
+1\right]\\
\nonumber&&\:-\frac{3e^4}{4\pi g}
\bigg[2\left(\frac{1}{\epsilon}-\gamma-2
-\ln\frac{m_{v}^2}{4\pi\mu^2}\right)
\:+\:\frac{1}{\left(1-\frac{3e^2}{g}\right)^2}
\:-\:\frac{2}{\left(1-\frac{3e^2}{g}\right)^3}
\ln\frac{3e^2}{g}\bigg]\\
&&\:-\frac{9e^6}{4\pi g^2}
\bigg[-\frac{7}{\left(1-\frac{3e^2}{g}\right)^2}
\:+\:\frac{2}{\left(1-\frac{3e^2}{g}\right)^3}
\ln\frac{3e^2}{g}\bigg]
\:+\frac{27e^8}{4\pi g^3}
\frac{6}{\left(1-\frac{3e^2}{g}\right)^3}\ln\frac{3e^2}{g}\,,\\
Z_{R,v}&=&1\:+\:\frac{e^2}{4\pi m_{v}^2}
\bigg[\frac{7m_{v}^4-m^2m_{v}^2}{(m^2-m_{v}^2)^2}
\:+\:2m_{v}^4\frac{m^2+2m_{v}^2}{(m^2-m_{v}^2)^3}
\ln\frac{3e^2}{g}\,\bigg]
\end{eqnarray}
in the first renormalization scheme, and for the pole mass
\begin{eqnarray}
\nonumber
m_A^2&=&m_{v}^2\:-\:\frac{g}{12\pi}
\Bigg[\ln\frac{3e^2}{g}\,+\,
\frac{2 \left(1-6\frac{e^2}{g}\right)^2}{\sqrt{12\frac{e^2}{g}-1}}
\arctan\sqrt{12\frac{e^2}{g}-1}\,\Bigg]\\
&&-\,\frac{e^2}{2\pi}\bigg[2\left(\frac{1}{\epsilon}
-\gamma+1-\ln\frac{m^2}{4\pi\mu^2}\right)
\,-\,\ln\frac{3e^2}{g}\bigg]
\,-\frac{3e^4}{2\pi g}\,
\bigg[\frac{1}{\epsilon}-\gamma-2-\ln\frac{m_{v}^2}{4\pi\mu^2}\bigg],\\
\nonumber  Z_{v}&=&1\,
+\,\frac{g}{12\pi m_{v}^2}
\Bigg[\ln\frac{3e^2}{g}
\,-\,2\,\frac{1-21\frac{e^2}{g}+108\frac{e^4}{g^2}-108\frac{e^6}{g^3}}
{\left(12\frac{e^2}{g}-1\right)^{\frac{3}{2}}}\,
\arctan\sqrt{12\frac{e^2}{g}-1}\,\Bigg]\\
&&+\,\frac{e^2}{4\pi m_{v}^2}
\Bigg[-\ln\frac{3e^2}{g}
-2\frac{\left(1-6\frac{e^2}{g}\right)^2}
{12\frac{e^2}{g}-1}\Bigg]
\end{eqnarray}
in the second scheme, where again we assume $g \leq 12 e^2$.

Using the correct prescription for taking the limit $\xi \to 0$, one
finds that the longitudinal part $\Pi_{2,\xi} (p^2)$ goes over to the
result of the U-gauge.

In one-loop order the mixing between the $A_{\mu}$ and $\phi_{2}$
reappears.  We do not display our results for the $\phi_{2}$ propagator
and the $A_{\mu}-\phi_{2}$ mixing, because Slavnov-Taylor identities
guarantee that contributions from the longitudinal gauge field, the
$\phi_{2}$ field and the ghosts cancel in physical amplitudes
\cite{LeeZJ2,LeeZJ4}.

We also calculated the ghost propagator in the R$_{\xi}$-gauge, but do
not display the result here.  It develops a pole at a non-vanishing
ghost mass.  In the limit $\xi \to 0$ the ghost mass goes to infinity,
as it should do \cite{Abers,LeeZJ4}, and the ghost propagator goes over
into the static one of the U-gauge.

%
\subsection{Field expectation value}

The vacuum expectation value of the Higgs field gets contributions from
one-loop diagrams. In the R$_{\xi}$-gauge one gets
\begin{eqnarray}
\nonumber v_{R_{\xi}}&=&v\:+\:\raisebox{-25\unitlength}{%
\begin{fmfgraph*}(50,50)
\fmfbottom{u}
\fmftop{o}
\fmf{plain,label=$p=0$}{u,v}
\fmf{phantom}{u,v,vi,o}
\fmf{plain,left=1}{v,o}
\fmf{plain,left=1}{o,v}
\fmfdot{v}
\end{fmfgraph*}}%
\:+\:\raisebox{-25\unitlength}{%
\begin{fmfgraph*}(50,50)
\fmfbottom{u}
\fmftop{o}
\fmf{plain,label=$p=0$}{u,v}
\fmf{phantom}{u,v,vi,o}
\fmf{dashes,left=1}{v,o}
\fmf{dashes,left=1}{o,v}
\fmfdot{v}
\end{fmfgraph*}}%
\:+\:\raisebox{-25\unitlength}{%
\begin{fmfgraph*}(50,50)
\fmfbottom{u}
\fmftop{o}
\fmf{plain,label=$p=0$}{u,v}
\fmf{phantom}{u,v,vi,o}
\fmf{dots_arrow,left=1}{v,o}
\fmf{dots,left=1}{o,v}
\fmfdot{v}
\end{fmfgraph*}}%
\:+\:\raisebox{-25\unitlength}{%
\begin{fmfgraph*}(50,50)
\fmfbottom{u}
\fmftop{o}
\fmf{plain,label=$p=0$}{u,v}
\fmf{phantom}{u,v,vi,o}
\fmf{boson,left=1}{v,o}
\fmf{boson,left=1}{o,v}
\fmfdot{v}
\end{fmfgraph*}}\\
\nonumber&=&\mu^{-\epsilon}v\Bigg\{1\:+\:\frac{g}{4\pi m^2}
\bigg[-\frac{2}{3}\left(\frac{1}{\epsilon}-\gamma
-\ln\frac{m^2}{4\pi\mu^2}\right)
\:+\:\frac{1}{6}\ln\frac{3e^2}{g}-\frac{1}{6}\ln\xi\,\bigg]\\
&&\:-\:\frac{e^2}{4\pi m^2}\left(\frac{1}{\epsilon}-\gamma-2
-\ln\frac{m_{v}^2}{4\pi\mu^2}\right)\,\Bigg\}.
\end{eqnarray}
In the proper limit it approaches the result of the U-gauge
\begin{eqnarray}
\nonumber v_{U}&=&v\:+\:\raisebox{-25\unitlength}{%
\begin{fmfgraph*}(50,50)
\fmfbottom{u}
\fmftop{o}
\fmf{plain,label=$p=0$}{u,v}
\fmf{phantom}{u,v,vi,o}
\fmf{plain,left=1}{v,o}
\fmf{plain,left=1}{o,v}
\fmfdot{v}
\end{fmfgraph*}}%
\;+\;
\raisebox{-25\unitlength}{%
\begin{fmfgraph*}(50,50)
\fmfbottom{u}
\fmftop{o}
\fmf{plain,label=$p=0$}{u,v}
\fmf{phantom}{u,v,vi,o}
\fmf{boson,left=1}{v,o}
\fmf{boson,left=1}{o,v}
\fmfdot{v}
\end{fmfgraph*}}\\
\nonumber&=&\mu^{-\epsilon}v\Bigg\{1\:-\:\frac{g}{8\pi m^2}
\bigg[\frac{1}{\epsilon}-\gamma-\ln\frac{m^2}{4\pi\mu^2}\bigg]
\:-\:\frac{e^2}{4\pi m^2}\bigg[\frac{1}{\epsilon}-\gamma-2
-\ln\frac{m_{v}^2}{4\pi\mu^2}\bigg]\Bigg\}.\\
\end{eqnarray}

The expressions for the field expectation value can be renormalized by
multiplication with an appropriate scalar field renormalization factor
$Z^{-1/2}$ and expressing the bare couplings and masses by their
renormalized counterparts. This would require the calculation of
three-point vertices in the one-loop approximation.

The vacuum expectation value of the scalar field is not independent of
the gauge parameter $\xi$, even after it is renormalized. This is not
unexpected \cite{Appel,Jacki,Dolan,Aitch}, since it is an off-shell
quantity.

%
\section{Conclusion}

The two-dimensional abelian Higgs model has been studied in the
R$_{\xi}$-gauge and in the unitary gauge in the framework of
dimensional regularization, where $D=2-2\epsilon$. The propagators and
field expectation values have been calculated on the one-loop level. 
An apparent discrepancy between the two gauges has been resolved, and
it has been shown that the results in the unitary gauge can be obtained
from those of the R$_{\xi}$-gauge by taking the limit $\xi \to 0$
before removing the dimensional regularization via $\epsilon \to 0$.
The resulting renormalized propagators are finite off-shell. The unitary
gauge appears to be suitable for the calculation of physical quantities.

It is, however, not possible to obtain the results for off-shell
($\xi$-dependent) quantities in the unitary gauge by taking the limit
$\xi \to 0$ of the final renormalized results in the R$_{\xi}$-gauge
after the regularization has been removed. For physical
$\xi$-independent quantities this reservation does not apply.

\end{fmffile}

%
\end{document}